\documentclass[twocolumn,preprintnumbers,superscriptaddress,amsmath,amssymb]{revtex4} 
\usepackage{graphicx}
\usepackage{dcolumn}
\usepackage{bm}
\usepackage{amssymb}
\usepackage{amssymb}
\usepackage{verbatim}
\usepackage{color}
\def\>{\rangle}

\begin{document}

\title{Dynamic Brillouin cooling for continuous optomechanical systems}
\author{Changlong Zhu}
\affiliation{Max Planck Institute for the Science of Light, Staudtstr. 2, 91058 Erlangen, Germany}
\affiliation{Department of Physics, University of Erlangen-Nuremberg, Staudtstr. 7, 91058 Erlangen, Germany}
\author{Birgit Stiller}\email{birgit.stiller@mpl.mpg.de}
\affiliation{Max Planck Institute for the Science of Light, Staudtstr. 2, 91058 Erlangen, Germany}
\affiliation{Department of Physics, University of Erlangen-Nuremberg, Staudtstr. 7, 91058 Erlangen, Germany}

\date{\today}

\begin{abstract}
In general, ground state cooling using optomechanical
interaction is realized in the regime where optical dissipation is higher
than mechanical dissipation. Here, we demonstrate that optomechanical
ground state cooling in a continuous optomechanical system is possible
by using backward Brillouin scattering while mechanical dissipation
exceeds optical dissipation which is the common case in optical waveguides.
The cooling is achieved in an anti-Stokes backward Brillouin process
by modulating the intensity of the optomechanical coupling via a pulsed
pump to suppress heating processes in the strong coupling regime.
With such dynamic modulation, a cooling factor with several orders of magnitude can be realized,
which breaks the steady-state cooling limit. This modulation scheme can also be
applied to Brillouin cooling generated by forward intermodal Brillouin scattering.

\end{abstract}


\maketitle

{\it Introduction}.---Cooling a mechanical oscillator to its ground
state by overcoming the effects of thermal environment has always
attracted great interests, as it offers attractive opportunities
for various topics including high precision metrology~\cite{LaHaye,Teufel1,Purdy},
quantum information processing~\cite{Palomaki,Riedinger,Ockeloen-Korppi},
and the exploration of classical-and-quantum limit of macroscopic
objects~\cite{Marshall,Pikovski,Arndt}. Preparing a single mode
mechanical oscillator into its quantum ground state has been
experimentally realized in cavity optomechanical systems by
utilizing methods in combination with optomechanical radiation
pressure interactions~\cite{Teufel2,Chan,Liu,Aspelmeyer}.
Apart from optomechanical radiation pressure interaction,
optoacoustic Brillouin interaction induced by electrostrictive effects~\cite{Boyd,Agrawal}
provides a potential mechanical cooling method for cavity optomechanical
systems~\cite{Matthew,Matthew2,Bahl,Chunhua,Chunhua2,Enzian}
and continuous optomechanical systems~\cite{Chen,Otterstrom}.
For both kinds of systems, nevertheless, Brillouin cooling has
so far only been studied for optical forward scattering.
In particular for a continuous waveguide system, forward Brillouin
cooling has been investigated theoretically~\cite{Chen} as
well as experimentally~\cite{Otterstrom}. However, Brillouin cooling
generated by backward scattering and cooling traveling-wave acoustic
phonons close to or even well into quantum ground state in continuous
optomechanical systems are still open questions.

Recently, a variety of integrated optomechanical waveguides at small size
scales ($\sim$cm or mm length) were realized in experiment~\cite{Eggleton}.
These short Brillouin-active waveguides with high Brillouin gain
allow the coherent light-sound interaction in a small regime, which
enables the dynamical control of photonic-phononic interaction through a
pulsed laser such as coherent photonic-phononic memory~\cite{Merklein}.
In addition, it has been theoretically predicted that the strong
coupling regime of the anti-Stokes Brillouin interaction can be
accessible in highly nonlinear waveguides~\cite{Laer1,Huy}.
The strong optomechanical interaction permits state swapping
between photons and phonons~\cite{Verhagen,Naeini} which is one way to
achieve phonon cooling~\cite{Hensinger,TianLin,TianLin2,Xiaoting,Yongchun}.
As a consequence, a Brillouin cooling scheme that can beat the
phonon heating rate for continuous optomechanical systems coupled
with the environment is highly desirable by manipulating the
dynamics of photonic-phononic interaction in integrated waveguides.

In this work, we demonstrate that one can achieve a great cooling factor
via backward Brillouin scattering in continuous optomechanical systems
under the strong coupling regime. By periodically modulating the optomechanical
coupling strength through a pulsed laser, the heating generated by the state
swapping and thermal noise can be significantly suppressed, which enables
the phonon occupancy to reach an
\begin{figure}[h]
\centerline{
\includegraphics[width=8.6 cm, clip]{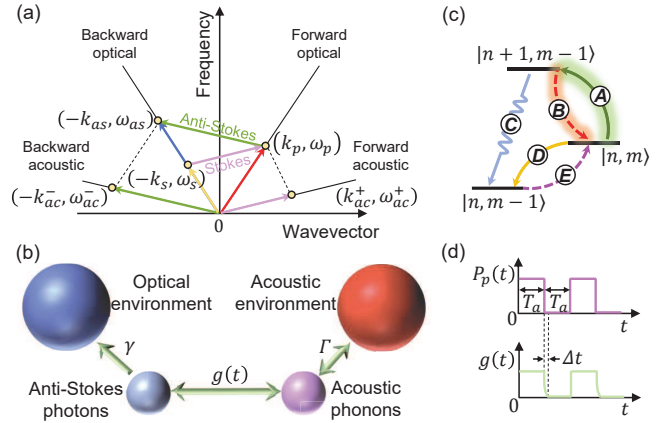}}
\caption{(color online) (a) Sketch of a typical dispersion diagram
of backward Brillouin scattering for both Stokes and anti-Stokes
processes. Subscripts of `$p$', `$as$', `$s$' correspond to optical
pump, anti-Stokes, Stokes fields and `$ac$' indicates acoustic field,
respectively. Superscripts `$-$' and `$+$' denote the
backward and forward direction separately.
(b) Schematic diagram of the linearized Brillouin anti-Stokes interaction.
(c) Level diagram of the Brillouin cooling in the strong coupling regime
where $|n,m\rangle$ denotes the state of $n$ anti-Stokes photons and $m$ acoustic phonons.
(d) Modulation of $g(t)$ via a pulsed laser $P_p(t)$ in short
Brillouin-active waveguides where $\Delta t\ll T_a \sim \pi/(2g(0))$.
}\label{Fig1-1}
\end{figure}
instantaneous-state cooling limit and
thereby breaks the fundamental limit of Brillouin cooling. We also
prove that this modulation scheme can be applied to Brillouin cooling
produced by forward Brillouin scattering and overcome the saturation
of the steady-state cooling limit.

In a typical Brillouin-active waveguide, the backward Brillouin
light-scattering is triply resonant where the Stokes
and anti-Stokes processes associate with counter-propagating
traveling-wave acoustic phonons, as show in Fig.~\ref{Fig1-1} (a).
This results in a natural dispersive symmetry breaking between
the Stokes and anti-Stokes processes~\cite{Kharel}.
It enables us to individually study the anti-Stokes process and
explore the optomechanical cooling, since the dynamics
of Stokes and anti-Stokes processes are independent
of each other. By applying an undepleted pump laser, this
triply resonant anti-Stokes process can be reduced
to a linearized optomechanical interaction between anti-Stokes
photons and acoustic phonons by considering an effective
pump-enhanced coupling strength $g(t)$~\cite{Laer1} which is modulated by the pump field.
Thus the anti-Stokes Brillouin interaction can be treated as
a beam-splitter-like interaction with the excitation (photon or phonon)
exchange between optical anti-Stokes and acoustic fields
at the coupling rate $g(t)$. This is illustrated in Fig.~\ref{Fig1-1} (b)
where $\gamma$ and $\Gamma$ denote optical and acoustic dissipation rates, respectively.
As the frequency of the optical anti-Stokes field is sufficiently
high, the anti-Stokes field sits its quantum ground state
and can be seen as equivalent to be coupled to an optical thermal
environment at effectively zero temperature. With
the excitation exchange, the optical anti-Stokes field constitutes
a source of essentially zero entropy for the acoustic field
and thus extracts the phonons out of the acoustic field.

When the pump power is strong enough, the system enters
the strong coupling regime which leads to a high fidelity
transfer of quantum states between the optical anti-Stokes
and acoustic fields, i.e., state swapping including swapping
heating and swapping cooling. We show the level diagram of
this linearized optomechanical interaction in Fig.~\ref{Fig1-1} (c).
Solid curves correspond to cooling processes including
swapping cooling ($A$), optical dissipation ($C$), and mechanical
dissipation ($D$) and dashed curves denote heating processes
containing swapping heating ($B$) and thermal heating ($E$).
Suppressing heating processes while enhancing cooling
processes in pursuit of an efficient phonon cooling rate is
the ultimate goal for Brillouin cooling. It should be noted
that the swapping cooling and heating processes dominate
alternately with a period $T_a$ ($\sim\pi/(2g)$) in the
strong coupling regime~\cite{TianLin}.

In the regime, where lights experience much lower dissipation than
phonons (typical waveguide Brillouin interaction), the phonon heating rate
induced by thermal noise (process $E$) exceeds the cooling
rate associated to optical dissipation (process $C$), which greatly limits
the cooling factor in the steady state. However, we can overcome
this limitation by dynamically tailoring the two swapping processes through exploiting
the coupling strength. For a Brillouin integrated waveguide~\cite{Eggleton},
if its length is short enough that the time $\Delta t$ consumed by lights
passing through the waveguide is far smaller than the evolution time
$T_a$ of each swapping process, the Brillouin optomechanical interaction
can be modulated by a pulsed pump laser, as shown in Fig.~\ref{Fig1-1} (d).
This allows a higher phonon cooling factor in the dynamic regime by
enhancing the swapping cooling process while suppressing the swapping
heating process and thus breaks the steady-state cooling limit.

To begin our discussion, we first analyze the dynamics of the mean
phonon number in the strong coupling regime. By considering an
undepleted constant CW pump laser, the dynamics of the linearized
anti-Stokes Brillouin interaction can be given by~\cite{Chen,Laer1}
\begin{eqnarray}\label{Dynamics of reduced Brillouin interaction}
\frac{\partial a_{as}}{\partial t} - \upsilon_{o}\frac{\partial a_{as}}{\partial z}
&=& -\frac{\gamma}{2}a_{as}- i g b_{ac} + \sqrt{\gamma}\xi_{as}, \nonumber\\
\frac{\partial b_{ac}}{\partial t} - \upsilon_{ac}\frac{\partial b_{ac}}{\partial z}
&=& -\frac{\Gamma}{2}b_{ac} - i g a_{as} + \sqrt{\Gamma}\xi_{ac},
\end{eqnarray}
where $a_{as}$ ($b_{ac}$) and $\upsilon_{o}$ ($\upsilon_{ac}$)
denote the envelope operator and group velocity of
the optical anti-Stokes (acoustic) field.
$\xi_{as}$ and $\xi_{ac}$ are the Langevin noise operators for
the optical anti-Stokes and acoustic fields.
$g = g_0\sqrt{\langle a_p^{\dagger}a_p\rangle}$
is the pump-enhanced coupling strength where $g_0$ indicates the
interaction strength between a single anti-Stokes photon
and a single phonon and $a_{p}$ represents pump envelope.
Without loss of generality, we take $g$ real and positive~\cite{Laer1}.
Actually, $a_{as}$ and $b_{ac}$ are modes with a continuous wavenumber and can
be expressed as $a_{as} = 1/\sqrt{2\pi} \int d k\ a_k e^{i (k-k_{as0}) z}$ and
$b_{ac} = 1/\sqrt{2\pi} \int d k\ b_k e^{i (k-k_{ac0}) z}$~\cite{Kharel,Sipe,Zoubi},
which are peaked around the carrier wave vector $k_{as0}$ (anti-Stokes wave)
and $k_{ac0}$ (acoustic wave), respectively.
Moving to momentum space by replacing $a_{as}$, $b_{ac}$,
$\xi_{as}$, $\xi_{ac}$, and $\partial/\partial z$ with
$a$, $b$, $\xi_1$, $\xi_2$, and $i k$, Eq.~(\ref{Dynamics of reduced Brillouin interaction})
can be re-expressed as
\begin{eqnarray}\label{Dynamical equation in momentum space}
\dot{a} &=& \left(-\gamma/2 + i \Delta_1 \right)a
- i g b + \sqrt{\gamma}\xi_1, \nonumber\\
\dot{b} &=& \left( -\Gamma/2 + i\Delta_2 \right)b
- i g a + \sqrt{\Gamma}\xi_2,
\end{eqnarray}
where $a(t,k)$ ($b(t,k)$) is the inverse Fourier transform
of the envelope operator $a_{as}(t,z)$ ($b_{ac}(t,z)$) and denotes
the annihilation operator for the $k$th photon (phonon) mode,
where the subscript $k$ for $a(t,k)$ ($b(t,k)$) has been
omitted for simplicity. $\Delta_1=k\upsilon_{as}$ and
$\Delta_2 = k\upsilon_{ac}$ induced by the wavenumber $k$
are the frequency shifts for the anti-Stokes photons and acoustic phonons,
where $\Delta_1=\Delta_2=0$ corresponds to the case when
the anti-Stokes optical mode and the acoustic mode are
phase-matched with the pump mode.
The Langevin noise terms $\xi_1$ and $\xi_2$ which are the inverse
Fourier transform of $\xi_{as}$ and $\xi_{ac}$ obey the correlations~\cite{Boyd2,Rakich}
$\langle\xi_i\rangle=0$, $\langle \xi^{\dagger}_1(t_1) \xi_1(t_2)\rangle=0$,
and $\langle \xi^{\dagger}_2(t_1) \xi_2(t_2)\rangle = n_{th}\delta(t_1-t_2)$,
where $n_{th}=1/(e^{\hbar\omega_{ac}/k_B T_m}-1)$ is the
thermal phonon occupation with frequency $\omega_{ac}$ at the
environment temperature $T_m$.

\begin{figure}[h]
\centerline{
\includegraphics[width=8.6 cm, clip]{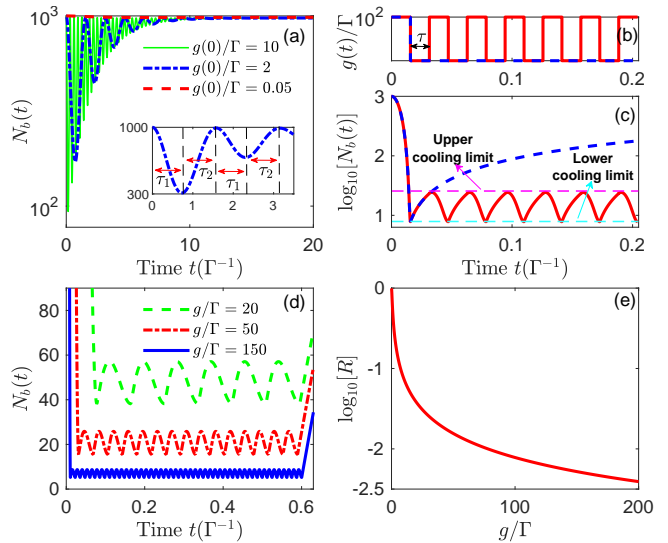}}
\caption{(color online) (a) Time evolution of mean phonon number
$N_b(t)$ for $g/\Gamma=0.05, 2, 10$ where the inset shows the
initial transient process of $N_b(t)$ for the case $g/\Gamma=2$.
(b) shows the dynamical modulation of $g(t)$
and (c) describes the corresponding time evolution of $N_b(t)$,
where blue dashed (red solid) curves
denote the single pulse (periodic pulse) modulation.
(d) Time evolution of $N_b(t)$ under pulsed modulation
of the coupling strength with different intensities.
(e) Brillouin cooling factor $R$ versus the coupling strength.
Other parameters are $\gamma/\Gamma=0.01$, $\Delta_1/\Gamma=0.3$,
$\Delta_2/\Gamma=3\times 10^{-5}$, and $n_{th}=1000$.
}\label{Fig2-1}
\end{figure}

We focus on the strong coupling regime ($\gamma,\Gamma\ll g$) and
consider that optical and acoustic frequency shifts are within the
linewidth of the acoustic mode ($\Delta_{1,2}<\Gamma$). In addition,
for the backward Brillouin scattering in a typical
waveguide, the mechanical dissipation is generally
far larger than the optical dissipation ($\Gamma\gg\gamma$)
and $\Delta_2\ll\Delta_1$ when $k\neq 0$ because of the slow
acoustic group velocity ($\upsilon_{ac}\ll\upsilon_{o}$).
Therefore, combining the Langevin equation described by
Eq.~(\ref{Dynamical equation in momentum space})
with noise correlations, a set of differential equations
for second-order moments $N_a=\langle a^{\dagger}a\rangle, N_b=\langle
b^{\dagger}b\rangle, \langle a^{\dagger}b\rangle$ can
be obtained (see the Supplemental Material) where
$N_b$ and $N_a$ correspond to the mean phonon and photon numbers.
By solving these differential equations, the analytical expression
of the time evolution of the mean phonon number can be given by
\begin{eqnarray}\label{Analytical solution}
N_b &=& N_{b,1} + N_{b,2}, \nonumber\\
N_{b,1} & \simeq & -n_{th}\frac{ 2(\Gamma-\gamma)g^2-\gamma(\Delta_1^2+\gamma\Gamma) }
{ (\gamma+\Gamma)\Omega^2 }
e^{-\frac{\gamma+\Gamma}{2}t} \nonumber\\
&&+ n_{th}\frac{\Gamma}{\Gamma+\gamma}, \nonumber\\
N_{b,2} & \simeq & n_{th}\frac{ 2g^2-\gamma(\gamma+\Gamma)/4 }{\Omega^2}
e^{-\frac{\gamma+\Gamma}{2}t}\cos(\Omega t) \nonumber\\
&&+ n_{th} \frac{ \gamma g -\gamma(\gamma+\Gamma)^2/(16g) }{\Omega^2}
e^{-\frac{\gamma+\Gamma}{2}t}\sin(\Omega t),
\end{eqnarray}
where $\Omega\approx\sqrt{4g^2 + 2\Delta_1^2-(\Gamma-\gamma)^2/4}$ .
We note that the phonon occupancy experiences a Rabi oscillation with
an exponentially decaying envelope and  can be divided into two parts
$N_{b,1}$ and $N_{b,2}$.
$N_{b,1}$ does not experience oscillation and tends to the steady-state cooling
limit ($\sim n_{th}\Gamma/(\Gamma+\gamma)$) with the exponentially
decaying rate $(\gamma+\Gamma)/2$, which implies
effects of optical and mechanical dissipations on the phonon cooling.
$N_{b,2}$ exhibits a Rabi oscillation with period $T\sim \pi/g$, which reveals the
energy transfer between photons and phonons in the strong coupling regime.
As depicted in Fig.~\ref{Fig1-1} (c), we see that two phonon cooling routes exist
including the route $A\rightarrow C$ and route $D$ where the first route
($A\rightarrow C$) is constrained by the optical dissipation process $C$.
In the strong coupling regime while the energy exchanging rate between
photons and phonons exceeds the optical dissipation rate,
this constrain in the first cooling route will cause
a saturation of phonon cooling for a higher coupling strength. This
is the reason that the phonon cooling speed, i.e., the exponentially
decaying envelope in Eq.~(\ref{Analytical solution}),
and the steady-state cooling limit are independent of the coupling strength
$g$ in the strong coupling regime. We present simulation results of time
evolution of the phonon occupancy $N_b(t)$ with different coupling strength $g$
in Fig.~\ref{Fig2-1} (a). It confirms that the strong optomechanical coupling
does not generate a faster cooling speed and a significant lower steady-state
cooling limit comparing with weak coupling when the optical dissipation
is far smaller than the mechanical dissipation.

{\it Breaking the steady-state cooling limit}.---Although
the strong optomechanical interaction does not significantly contribute
to the steady-state cooling limit, it results in a Rabi oscillation
for the phonon occupancy and thus leads the minimum phonon occupancy to
be far smaller than the steady-state cooling limit, as shown in Fig.~\ref{Fig2-1} (a).
The system transfers from state $|n,m\rangle$ to state $|n,m-1\rangle$
by extracting phonons out of the acoustic field during the swapping-cooling-dominant
time period ($\tau_1$). Analogously, it transfers from $|n,m\rangle$ to
$|n,m+1\rangle$ by generating phonons during the swapping-heating-dominant
time period ($\tau_2$). These two time periods alternate with a cycle
$\tau_1=\tau_2=T/2$, as shown in the inset of Fig.~\ref{Fig2-1} (a).
To break the steady-state cooling limit and obtain a significant cooling rate,
here we dynamically modulate the coupling strength $g(t)$
through a pulsed pump laser to permit the swapping cooling process
and suppress the swapping heating process.
We consider a short enough Brillouin-active waveguide, i.e.,
$L\ll \upsilon_{o}T/2$, where light fields will quickly pass through
the waveguide and thus a pump laser can generate a coupling
strength with pulsed-shape by switching on and off the pump.
We switch on the pump laser during the swapping-cooling-dominant time
period ($\tau_1$) to strengthen phonon absorption and switch off the pump
during the swapping-heating-dominant time period ($\tau_2$) to halt
the reversible Rabi oscillation and thus suppress the swapping heating process.
We illustrate the modulation scheme of the pulsed coupling strength
and the corresponding time evolution of the phonon occupancy in Figs.~\ref{Fig2-1} (b)
and (c), respectively.
Since the phonon occupancy reaches the minimum value at the end
of the first half Rabi oscillation, we switch off the pump abruptly
at this time to prevent the energy from transferring back to phonons.
During the pump switch-off time period, the phonons are only driven
by the thermal environment, thus the phonon occupancy
increases with the exponential growing rate $\Gamma/2$. After the
optical fields passe through the waveguide, i.e., optical
fields are initialized to the vacuum state, we switch on the
pump laser to excite the swapping cooling process to absorb phonons
and prevent the phonon occupancy from increasing continuously.
By periodically modulating the coupling strength to initialize
optical fields regularly, we can continuously suppress the
swapping and thermal heating processes to keep a low phonon occupancy with
a small-amplitude fluctuation and thus break the steady-state
cooling limit, as shown by red solid curves in Figs.~\ref{Fig2-1} (b) and (c).
The instantaneous-state cooling limit, i.e., the lower cooling limit
in Fig.~\ref{Fig2-1} (c), can be approximately expressed as
$N_b^{\rm{ins}}\approx\pi\Gamma n_{th}/(4g)$ which reduces the Brillouin
steady-state cooling limit by a factor of $\pi\Gamma/(4g)$.
The upper cooling limit in Fig.~\ref{Fig2-1} (c) can be approximately expressed as
$N_b^{\rm{upp}}\approx(1+\pi)\Gamma n_{th}/(4g)$.
We know that in cavity optomechanical systems in the weak coupling regime,
the cooling limit in the resolved-sideband regime is mainly dependent on
the effective coupling strength $G$ and the optical dissipation
rate $\gamma$~\cite{{I-Wilson-Rae,F-Marquardt,C-Genes}}, i.e., $\sim n_{th}G^2/\gamma$.
Here, as we consider the strong coupling regime and periodically evacuate
the photons, the instantaneous-state cooling limit is decided by the ratio
between the mechanical dissipation rate and the effective coupling strength.

In fact, the small-amplitude fluctuation around the instantaneous-state
cooling limit is induced by the pulsed modulation of the coupling strength
and the strong optomechanical interaction, which can be optimized by tuning the
pump switch-off time $\tau$. In Fig.~\ref{Fig2-1} (d), we show the time
evolution of the phonon occupancy with periodical modulation of the
coupling strength while the pump switch-off time is $\tau=0.05T$, where
$g$ is the coupling strength during the pump switch-on time periods.
We also present the Brillouin cooling factor $R$, which is the ratio of the
instantaneous cooling limit to the initial phonon occupancy,
in Fig.~\ref{Fig2-1} (e).
It indicates that a great Brillouin cooling factor, which reduces the
steady-state cooling limit ($R\approx1$) by several orders of magnitude,
can be achieved through pulsed modulation of the optomechanical interaction,
while photons experience lower damping than phonons.
Moreover, this modulation scheme is switchable, i.e., the system will reach
the instantaneous-state or steady-state cooling limits by turning on or off
the modulation (see the Supplemental Material).

\begin{figure}[h]
\centerline{
\includegraphics[width=8.6 cm, clip]{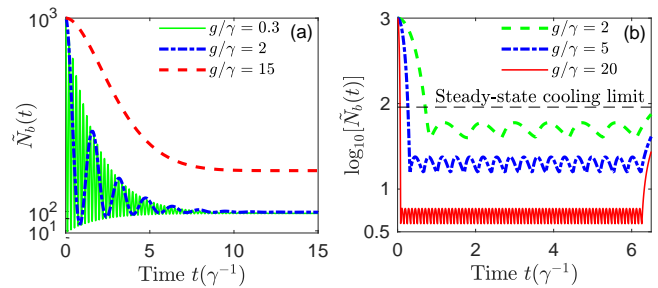}}
\caption{(color online) Dynamical Brillouin cooling via
forward anti-Stokes intermodal scattering.
(a) Time evolution of
$\tilde{N}_b(t)$ for $g/\gamma=0.3, 2, 15$.
(b) Dynamical cooling via pulsed modulation of the coupling intensity.
Other parameters are $\Gamma/\gamma=0.1$, $\Delta_1/\gamma=0.05$,
$\Delta_2/\gamma=5\times 10^{-6}$, and $n_{th}=1000$.
}\label{Fig3-1}
\end{figure}

In addition, the above modulation scheme can also be applied to
optomechanical cooling generated by forward anti-Stokes intermodal Brillouin
scattering in continuous optomechanical systems~\cite{Otterstrom}, where lights experience
higher damping than phonons. Actually, like the
backward scattering case stated above, the phonon occupancy $\tilde{N}_b(t)$
corresponding to the forward intermodal Brillouin scattering
exhibits a Rabi oscillation in the strong coupling regime, which enables lower
phonon occupancy at some instantaneous states, as shown in Fig.~\ref{Fig3-1} (a).
Here, we choose the ratio between mechanical and optical
dissipations according to the parameters measured in experiment~\cite{Otterstrom}.
With a pulsed modulation of the optomechanical interaction to suppress heating processes,
the phonon occupancy can be continuously maintained in a lower occupation with
a small-amplitude fluctuation, which breaks the steady-state cooling limit,
as shown in Fig.~\ref{Fig3-1} (b).

{\it Conclusion}.---We have shown that by periodically modulating
the Brillouin interaction with a pulsed pump in the strong coupling
regime, we can stimulate the swapping cooling process while suppressing
the swapping heating process and thus obtain a significant Brillouin
cooling factor with several orders of magnitude. It proves that cooling
traveling-wave phonons into quantum ground state by utilizing backward
Brillouin scattering is possible in continuous optomechanical systems
while mechanical dissipation exceeds optical dissipation.
Our scheme can also be applied to Brillouin cooling produced by
forward intermodal Brillouin scattering and break the steady-state cooling limit.
Moreover, this pulsed modulation scheme can be switchable.
In addition, distinct from other pulsed cooling
schemes that use complicated control methods~\cite{Xiaoting,Machnes} or dynamic
dissipative cooling by exploiting the optical dissipation~\cite{Yongchun}
in cavity optomechanical systems, the simplicity and convenience of our
method is achieved by simply controlling the pump pulse which makes the dynamic cooling
scheme an effective experimental tool for quantum optomechanics.
It should be pointed out that even though there is a small-amplitude fluctuation
of the phonon occupancy around the instantaneous-state cooling limit,
the Brillouin cooling limit can be viewed as stable in the sense of
time averaging while the time scale is larger than the Rabi oscillation
cycle $T$~\cite{Yongchun}. This work opens the way for the exploration
of quantum phenomena in continuous optomechanical systems through
backward Brillouin scattering. The dynamical control of optomechanical
interaction also provides a new way to study the quantum technologies,
ranging from mechanical quantum states generation, quantum information
processing, and high-precise measurement.

\begin{acknowledgements}
The authors would like to acknowledge very useful discussions with
Christian Wolff, Claudiu Genes, Florian Marquardt, and Yu-xi Liu.
This work is supported by the Max-Planck-Society through the independent
Max Planck Research Groups Scheme.
\end{acknowledgements}

\appendix
\section{Motion equation of linearized Brillouin interaction}\label{S1}
The dynamics of the anti-Stokes Brillouin backward scattering in a typical waveguide
can be given by
\begin{eqnarray}\label{Motion equation of complete Brillouin}
\frac{\partial a_p}{\partial t} + \upsilon_o\frac{\partial a_p}{\partial z} &=&
-\frac{\gamma}{2}a_p - i g_0 a_{as}b_{ac}^{\dagger}+\sqrt{\gamma}\xi_p, \nonumber\\
\frac{\partial a_{as}}{\partial t} - \upsilon_o\frac{\partial a_{as}}{\partial z} &=&
-\frac{\gamma}{2}a_{as} - i g_0a_p b_{ac} + \sqrt{\gamma}\xi_{as}, \nonumber\\
\frac{\partial b_{ac}}{\partial t} - \upsilon_b \frac{\partial b_{ac}}{\partial z}&=&
-\frac{\Gamma}{2}b_{ac} - i g_0 a_p^{\dagger}a_{as} + \sqrt{\Gamma}\xi_{ac},
\end{eqnarray}
where $a_p$, $a_{as}$, and $b_{ac}$ denote the envelope operators of the pump
field, anti-Stokes field, and acoustic field at their respective carrier frequencies
$\omega_p$, $\omega_{as}$, and $\omega_{ac}$. $\upsilon_o$ and $\upsilon_{ac}$ represent
the group velocities of the optical and acoustic fields. $\gamma$ and $\Gamma$ are
the dissipation rates of the optical and acoustic fields. $g_0$ is the traveling-wave
vacuum coupling rate which quantifies the interaction intensity between a single phonon
and a single photon\cite{Laer1}. Without loss of generality, we take $g_0$ real and
positive in our discussion. $\Delta_1=k\upsilon_{o}$ and $\Delta_2=k\upsilon_{ac}$
are the frequency shifts for the anti-Stokes photons and acoustic phonons which are
induced by the wavenumber $k$, where $\Delta_1=\Delta_2=0$ corresponds to the case when
the anti-Stokes optical mode and the acoustic mode are phase-matched with the pump mode.
$\xi_p$, $\xi_{as}$, and $\xi_{ac}$ correspond to the Langevin noises of the pump field,
anti-Stokes field, and acoustic field, which obey the following mean and correlation
\begin{eqnarray}
\langle \xi_p(t,z)\rangle &=& 0, \nonumber\\
\langle \xi_{as}(t,z) \rangle &=& 0, \nonumber\\
\langle \xi_{ac}(t,z) \rangle &=& 0, \nonumber\\
\langle \xi_p^{\dagger} (t_1,z_1)\xi_p(t_2,z_2)\rangle &=& 0, \nonumber\\
\langle \xi_{as}^{\dagger}(t_1,z_1)\xi_{as}(t_2,z_2)\rangle &=& 0, \nonumber\\
\langle \xi_{ac}^{\dagger}(t_1,z_1)\xi_{ac}(t_2,z_2)\rangle &=& n_{th}\delta(t_1-t_2)\delta(z_1-z_2),\nonumber\\
\end{eqnarray}
where $n_{th}=1/(e^{\hbar\omega_{ac}/k_B T_m}-1)$ is the thermal phonon occupation
at the environment temperature $T_m$. By applying an undepleted pump field,
the triply resonant optomechanical interaction can be reduced to a linearized optomechanical
interaction between anti-Stokes field and acoustic field with a pump-enhanced
coupling strength. Thus Eq.~(\ref{Motion equation of complete Brillouin}) can be reduced to
\begin{eqnarray}\label{Motion equation of linearized Brillouin}
\frac{\partial a_{as}}{\partial t} - \upsilon_o\frac{\partial a_{as}}{\partial z} &=&
-\frac{\gamma}{2}a_{as} - i g b_{ac} + \sqrt{\gamma}\xi_{as}, \nonumber\\
\frac{\partial b_{ac}}{\partial t} - \upsilon_{ac} \frac{\partial b_{ac}}{\partial z}&=&
-\frac{\Gamma}{2}b_{ac} - i g a_{as} + \sqrt{\Gamma}\xi_{ac},
\end{eqnarray}
where $g=g_0\sqrt{\langle a_p^{\dagger}a_p\rangle}$ is the pump-enhanced spatial coupling rate.
In fact, the phonon-mode and photon-mode $b_{ac}$ and $a_{as}$ are envelope operators with a
continuous wavenumber and peaked around carrier wave vectors $k_{ac0}$ and $k_{as0}$, respectively,
which can be expressed as~\cite{Kharel,Sipe}
\begin{eqnarray}\label{Envelope operators with continuous wavenumber}
b_{ac}&=&\frac{1}{\sqrt{2\pi}} \int d k b_k e^{i(k-k_{ac0})z}, \nonumber\\
a_{as}&=&\frac{1}{\sqrt{2\pi}} \int d k a_k e^{i(k-k_{as0})z},
\end{eqnarray}
where $b_k$ ($a_k$) denotes the annihilation operator for the $k$th phonon (photon)
mode. Now we move to the momentum space by replacing $a_{as}$, $b_{ac}$, $\xi_{as}$,
$\xi_{ac}$, and $\partial/\partial z$ with $a$, $b$, $\xi_1$, $\xi_2$, and $i k$
in Eq.~(\ref{Motion equation of linearized Brillouin}) and obtain the motion equation
of the linearized Brillouin interaction which can be given by
\begin{eqnarray}\label{Motion equation of linearized Brillouin with reduced variables}
\frac{d a(k,t)}{d t} &=& (-\frac{\gamma}{2} + i\Delta_1)a - i g b + \sqrt{\gamma}\xi_1,  \nonumber\\
\frac{d b(k,t)}{d t} &=&(-\frac{\Gamma}{2} + i\Delta_2)b - i g a + \sqrt{\Gamma}\xi_2,
\end{eqnarray}
where the subscript $k$ of photon and phonon annihilation operators have been removed
for simplification and $\xi_1$ ($\xi_2$) is the inverse Fourier transform of Langevin
noise $\xi_{as}$ ($\xi_{ac}$).
In order to derive the properties of Langevin noise $\xi_1(k,t)$, we decouple the
optomechanical interaction between photons and phonons and assume that the anti-Stokes
field $a(k,t)$ is driven by the Langevin noise $\xi_1(k,t)$, thus the quantum
Langevin equation of $a(,t)$ can be given by
\begin{eqnarray}\label{Langevin equation of optical field driven by noise}
\dot{a} = (i\Delta_1-\frac{\gamma}{2}) a + \sqrt{\gamma}\xi_1,
\end{eqnarray}
where $\xi_1$ is a Gaussian random variable with zero mean, i.e., $\langle \xi_1(k,t)\rangle=0$,
and $\delta$ correlation
\begin{eqnarray}
\langle \xi_1^{\dagger}(k,t_1)\xi_1(k,t_2) \rangle = Q_1\delta(t_1-t_2).
\end{eqnarray}
The solution of Eq.~(\ref{Langevin equation of optical field driven by noise}) can
be expressed as
\begin{eqnarray}
a(k,t) = e^{(i\Delta_1-\gamma/2)t}\int_{-\infty}^{t}e^{-(i\Delta_1-\gamma/2)\tau}
\sqrt{\gamma}\xi_1(k,\tau)d\tau.
\end{eqnarray}
Thus the equal-time correlation of $a$ can be expressed as
\begin{eqnarray}
&&\langle a^{\dagger}(k,t)a(k,t)\rangle \nonumber\\ 
&=&\gamma e^{-\gamma t}\int_{-\infty}^{t}
e^{-(-i\Delta_1-\gamma/2)\tau_1}d\tau_1 \nonumber\\  
&&\times\int_{-\infty}^{t}
e^{-(i\Delta_1-\gamma/2)\tau_2} \langle \xi_1^{\dagger}(k,\tau_1)\xi_1(k,\tau_2)\rangle d\tau_2 \nonumber\\
&=&Q_1.
\end{eqnarray}
In addition, $\langle a^{\dagger}(k,t)a(k,t)$ corresponds to the thermal photon occupation $n_{th,o}$,
i.e., $Q=n_{th,o}$. Hence, the correlation relation of Langevin noise $\xi_1$ can be given by
\begin{eqnarray}\label{Properties of optical noise nonzero}
\langle \xi_1^{\dagger}(k,t_1)\xi_1(k,t_2) \rangle &=& n_{th,o}\delta(t_1-t_2), \nonumber\\
\langle \xi_1(k,t_1) \xi_1^{\dagger}(k,t_1) \rangle &=& (1+n_{th,o}) \delta(t_1-t_2).
\end{eqnarray}
Since the frequency of the anti-Stokes photons is high enough that the anti-Stokes
field sits the quantum ground state, the thermal photon occupancy is zero, i.e., $n_{th,o}=0$.
Therefore, the properties of Langevin noise $\xi_1$ can be expressed as
\begin{eqnarray}\label{Properties of optical noise}
\langle \xi_1(k,t)\rangle &=& 0, \nonumber\\
\langle \xi_1^{\dagger}(k,t_1)\xi_1(k,t_2) \rangle &=& 0, \nonumber\\
\langle \xi_1(k,t_1) \xi_1^{\dagger}(k,t_1) \rangle &=& \delta(t_1-t_2).
\end{eqnarray}
Similarly, the properties of Langevin noise $\xi_2$ corresponding to the
acoustic mode can be given by
\begin{eqnarray}\label{Properties of acoustic noise}
\langle \xi_2(k,t)\rangle &=& 0, \nonumber\\
\langle \xi_2^{\dagger}(k,t_1) \xi_2(k,t_2) &=& n_{th}\delta(t_1-t_2), \nonumber\\
\langle \xi_2(k,t_1) \xi_2^{\dagger}(k,t_2) &=& (1+n_{th})\delta(t_1-t_2).
\end{eqnarray}

\section{Dynamics of mean phonon number in the strong coupling regime}\label{S2}
\begin{figure}[h]
\centerline{
\includegraphics[width=8.6 cm,clip]{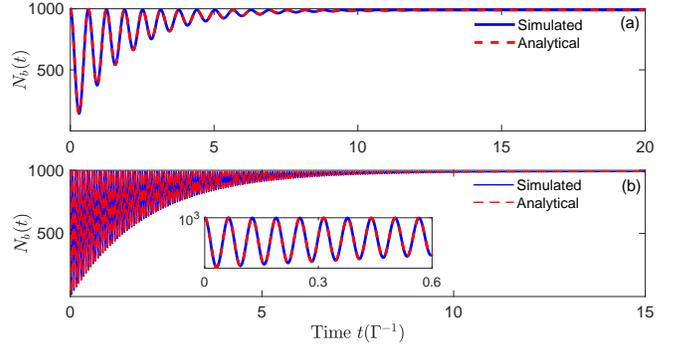}}
\caption{(Color online) Time evolution of phonon occupancy $N_b(t)$ in the strong coupling regime
for $g/\Gamma=5$ (a) and $g/\Gamma=50$ (b). The blue solid curves represent simulated results and
the red dashed curves correspond to the analytical expression described in
Eq.~(\ref{Reduced analytical solution of Nb}). The inset in (b) shows the initial transient process
of $N_b(t)$ for $g/\Gamma=50$.
Other parameters are $\gamma/\Gamma=0.01$, $\Delta_1/\Gamma=0.3$,
$\Delta_2/\Gamma=3\times 10^{-5}$, and $n_{th}=1000$.
}\label{sup_Fig1}
\end{figure}

Based on the motion equation of anti-Stokes photons and acoustic phonons described in
Eq.~(\ref{Motion equation of linearized Brillouin with reduced variables}), we have
\begin{eqnarray}\label{Motion equation of linearized Brillouin with reduced variables diagger}
\frac{d{a}^{\dagger}}{d t} &=& (-i\Delta_1-\frac{\gamma}{2}) a^{\dagger} + i g b^{\dagger} + \sqrt{\gamma}\xi_1^{\dagger}, \nonumber\\
\frac{d{b}^{\dagger}}{d t} &=& (-i\Delta_2-\frac{\Gamma}{2}) b^{\dagger} + i g a^{\dagger} + \sqrt{\Gamma}\xi_2^{\dagger}.
\end{eqnarray}
Combining with Eq.~(\ref{Motion equation of linearized Brillouin with reduced variables}),
these differential equations for the second-order moments $N_a=\langle a^\dagger a\rangle$,
$N_b=\langle b^{\dagger}b\rangle$, $\langle a^{\dagger}b\rangle$ can be written as
\begin{eqnarray}\label{Motion equation of second order moments with noise terms}
\frac{d N_a}{d t} &=& -\gamma N_a - i g ( \langle a^{\dagger} b\rangle - \langle a^{\dagger} b \rangle^* )
+ \sqrt{\gamma}( \langle \xi_1^{\dagger} a\rangle + \langle \xi_1^{\dagger} a\rangle^* ), \nonumber\\
\frac{d N_b}{d t} &=& -\Gamma N_b + i g ( \langle a^{\dagger} b\rangle - \langle a^{\dagger} b \rangle^* )
+ \sqrt{\Gamma} ( \langle \xi^{\dagger}_2 b \rangle + \langle \xi^{\dagger}_2 b \rangle^* ), \nonumber\\
\frac{d \langle a^{\dagger}b\rangle}{d t} &=& - \left (i(\Delta_1-\Delta_2)+\frac{\gamma+\Gamma}{2}\right )
\langle a^{\dagger}b\rangle - i g N_a + i g N_b  \nonumber\\
&&+ \sqrt{\gamma}\langle \xi_1^{\dagger} b \rangle + \sqrt{\Gamma} \langle a^{\dagger} \xi_2\rangle,
\end{eqnarray}
where $N_a$ and $N_b$ denote the mean photon and phonon numbers, respectively.
In order to evaluate the noise-related terms in Eq.~(\ref{Motion equation of second order moments with noise terms}),
we apply the following Fourier transform
\begin{eqnarray}
\bar{a}(\omega) &=& \int_{-\infty}^{\infty} a(t)e^{i\omega t} d t, \nonumber\\
\bar{b}(\omega) &=& \int_{-\infty}^{\infty} b(t) e^{i\omega t} d t, \nonumber\\
\bar{\xi}_1(\omega) &=& \int_{-\infty}^{\infty} \xi_1(t) e^{i\omega t} d t, \nonumber\\
\bar{\xi}_2(\omega) &=& \int_{-\infty}^{\infty} \xi_2(t) e^{i\omega t} d t,
\end{eqnarray}
to Eq.~(\ref{Motion equation of linearized Brillouin with reduced variables}) and obtain
\begin{eqnarray}\label{Motion equation of a and b in frequency space}
-i\omega\bar{a} &=& (i\Delta_1-\frac{\gamma}{2})\bar{a} - i g \bar{b} + \sqrt{\gamma}\bar{\xi}_1, \nonumber\\
-i\omega\bar{b} &=& (i\Delta_2-\frac{\Gamma}{2})\bar{b} - i g \bar{a} + \sqrt{\Gamma}\bar{\xi}_2,
\end{eqnarray}
where the correlation relations of noises $\bar{\xi}_{1,2}$ can be calculated
based on the correlation functions described in Eqs.~(\ref{Properties of optical noise})
and (\ref{Properties of acoustic noise}) and given by
\begin{eqnarray}\label{Noise properties in frequency space}
\langle \bar{\xi}_1^{\dagger}(\omega_1)\bar{\xi}_1(\omega_2)\rangle &=& 0, \nonumber\\
\langle \bar{\xi}_2^{\dagger}(\omega_1)\bar{\xi}_2(\omega_2)\rangle &=& 2\pi n_{th}\delta(\omega_1-\omega_2), \nonumber\\
\langle \bar{\xi}_1^{\dagger}(\omega_1)\bar{\xi}_2(\omega_2)\rangle &=& 0, \nonumber\\
\langle \bar{\xi}_1(\omega_1)\bar{\xi}_2^{\dagger}(\omega_2)\rangle &=& 0.
\end{eqnarray}
The solution of $\bar{a}(\omega), \bar{b}(\omega)$ can be expressed as
\begin{eqnarray}\label{Solution of a and b in frequency space}
\bar{a}(\omega) &=& -\frac{ \left( i(\omega+\Delta_2)-\frac{\Gamma}{2} \right)\sqrt{\gamma}\bar{\xi}_1(\omega)
+ i g \sqrt{\Gamma}\bar{\xi}_2(\omega) }
{ \Xi }, \nonumber\\
\bar{b}(\omega) &=& -\frac{ i g \sqrt{\gamma}\bar{\xi}_1(\omega) + \left( i( \omega+\Delta_1 )
-\frac{\gamma}{2} \right)\sqrt{\Gamma}\bar{\xi}_2(\omega) }
{ \Xi }, \nonumber\\
\end{eqnarray}
with
\begin{eqnarray}
\Xi &=& g^2+\frac{\gamma\Gamma}{4} - (\omega+\Delta_1)(\omega+\Delta_2) \nonumber\\
&&- i \left( \frac{\gamma}{2}(\omega+\Delta_2)
+ \frac{\Gamma}{2}(\omega+\Delta_1) \right).
\end{eqnarray}
By substituting Eqs.~(\ref{Noise properties in frequency space}) and
(\ref{Solution of a and b in frequency space}) into $\langle \xi_1^{\dagger}(t)a(t)\rangle$,
we have
\begin{eqnarray}
&&\langle \xi_1^{\dagger}(t)a(t)\rangle \nonumber\\ &=& \frac{1}{4\pi^2}\int_{-\infty}^{\infty}
e^{i\omega_1 t} d \omega_1 \int_{-\infty}^{\infty} e^{-i\omega_2 t}
\langle \bar{\xi}_1^{\dagger}(\omega_1)\bar{a}(\omega_2)\rangle d \omega_2 \nonumber\\
&=&-\frac{1}{4\pi^2} \int_{-\infty}^{\infty} e^{i\omega_1 t} d \omega_1
\int_{-\infty}^{\infty} e^{-i\omega_2 t} \times \nonumber\\
&&\langle \bar{\xi}_1^{\dagger}(\omega_1) \frac{ \left( i(\omega_2+\Delta_2)-\frac{\Gamma}{2}
\right)\sqrt{\gamma}\bar{\xi}_1(\omega_2) + i g \sqrt{\Gamma}\bar{\xi}_2(\omega_2) }
{\Xi} \rangle d \omega_2 \nonumber \\
&=&0.
\end{eqnarray}
Similarly, $\langle \xi_2^{\dagger}b\rangle$, $\langle \xi_1^{\dagger}b\rangle$, and
$\langle a^{\dagger}\xi_2\rangle$ can be calculated as follows
\begin{eqnarray}
&&\langle \xi^{\dagger}_2(t)b(t)\rangle \nonumber\\
&=&\frac{1}{4\pi^2}\int_{-\infty}^{\infty} e^{i\omega_1t} d \omega_1
\int_{-\infty}^{\infty} e^{-i\omega_2t}\langle \bar{\xi}_2^{\dagger}(\omega_1)\bar{b}(\omega_2)\rangle d\omega_2 \nonumber\\
&=&-\frac{\sqrt{\Gamma}n_{th}}{2\pi} \int_{-\infty}^{\infty}
\frac{ i(\omega+\Delta_1)-\frac{\gamma}{2} }
{ \Xi } d\omega,
\end{eqnarray}
\begin{eqnarray}
&&\langle \xi_1^{\dagger}(t)b(t)\rangle \nonumber\\
&=&\frac{1}{4\pi^2} \int_{-\infty}^{\infty} e^{i\omega_1t} d\omega_1
\int_{-\infty}^{\infty} e^{-i\omega_2t}
\langle \bar{\xi}_1^{\dagger}(\omega_1)\bar{b}(\omega_2)\rangle d\omega_2 \nonumber\\
&=& 0,
\end{eqnarray}
and
\begin{eqnarray}
&&\langle a^{\dagger}(t)\xi_2(t)\rangle \nonumber\\
&=&\frac{1}{4\pi^2} \int_{-\infty}^{\infty} e^{i\omega_1t} d\omega_1
\int_{-\infty}^{\infty} e^{-i\omega_2t}
\langle \bar{a}^{\dagger}(\omega_1) \bar{\xi}_2(\omega_2) \rangle d\omega_2 \nonumber \\
&=&\frac{i g n_{th}\sqrt{\Gamma}}{2\pi}
\int_{-\infty}^{\infty}\frac{1}
{ \Xi } d \omega.
\end{eqnarray}
Hence, the noise-related terms in Eq.~(\ref{Motion equation of second order moments with noise terms})
can be expressed as follows
\begin{eqnarray}
\langle \xi_1^{\dagger}a\rangle + \langle \xi_1^{\dagger}a\rangle^* &=& 0, \label{equal time correlation between xi1 and a} \\
\langle \xi_2^{\dagger}b\rangle + \langle \xi_2^{\dagger}b\rangle^*
&=&\frac{\sqrt{\Gamma}n_{th}}{2\pi}
\int_{-\infty}^{\infty} \\
&& \frac{ \gamma(g^2+\frac{\gamma\Gamma}{4}) + \Gamma (\omega+\Delta_1)^2 }
{ \Lambda_1 } d \omega, \nonumber\\
\sqrt{\gamma}\langle \xi_1^{\dagger}b\rangle + \sqrt{\Gamma}\langle a^{\dagger}\xi_2\rangle
&=& \frac{ i g n_{th}\Gamma }{2\pi}\int_{-\infty}^{\infty}
\frac{1}{ \Lambda_2 } d\omega, \label{equal time correlation xi1 b and xi2 a}
\end{eqnarray}
with
\begin{eqnarray}
\Lambda_1 &=& \left( g^2+\frac{\gamma\Gamma}{4}- ( \omega^2+(\Delta_1+\Delta_2)\omega +\Delta_1\Delta_2 \right)^2 \nonumber\\
&&+ \left(  \frac{\gamma+\Gamma}{2}\omega + \frac{\gamma\Delta_2+\Gamma\Delta_1}{2} \right)^2, \nonumber\\
\Lambda_2 &=& g^2+\frac{\gamma\Gamma}{4} - (\omega+\Delta_1)(\omega+\Delta_2) \nonumber\\
&&+i\left( \frac{\gamma}{2}(\omega+\Delta_2)+\frac{\Gamma}{2}(\omega+\Delta_1) \right).
\end{eqnarray}
Here, we consider the strong coupling regime, i.e., $g\gg \Gamma,\gamma$. Furthermore,
for the backward Brillouin scattering in a typical Brillouin-active waveguide,
the mechanical dissipation is far larger than the optical dissipation ($\Gamma\gg\gamma$) and the
optical group velocity is significantly faster than the mechanical group velocity
($\upsilon_{o}\gg\upsilon_{ac}$, i.e., $\Delta_1\gg\Delta_2$). We also consider that
the wavenumber induced frequency shifts are within the linewidth of the acoustic
mode ($\Delta_{1,2}<\Gamma$). Thus $\langle \xi_2^{\dagger}b\rangle + \langle \xi_2^{\dagger}b\rangle^*$
can be approximated as follows
\begin{eqnarray}\label{equal time correlation between xi2 and b 1}
&&\langle \xi_2^{\dagger}(t)b(t)\rangle + \langle \xi_2^{\dagger}(t)b(t)\rangle^* \nonumber\\
&\approx& \frac{\sqrt{\Gamma}n_{th}}{2\pi} \int_{-\infty}^{\infty}
\frac{ \gamma(g^2+\frac{\gamma\Gamma}{4})+\Gamma\omega^2 }
{ \left( g^2+\frac{\gamma\Gamma}{4}+\Delta_1\omega-\omega^2 \right)^2
+ \left( \frac{\gamma+\Gamma}{2} \right)^2\omega^2 } d\omega \nonumber\\
&\approx& \frac{\sqrt{\Gamma}n_{th}}{2\pi} \int_{-\infty}^{\infty}
\frac{ \gamma\alpha^2+\Gamma\omega^2 }{ (\alpha^2+\Delta_1\omega-\omega^2)^2 + \beta^2\omega^2 } d\omega,
\end{eqnarray}
where
\begin{eqnarray}
\alpha = \sqrt{ g^2+\frac{\gamma\Gamma}{4} }, \quad \beta=\frac{\gamma+\Gamma}{2}.
\end{eqnarray}
Eq.~(\ref{equal time correlation between xi2 and b 1}) can be further calculated as follows
\begin{eqnarray}\label{equal time correlation between xi2 and b 2}
&&\frac{\sqrt{\Gamma}n_{th}}{2\pi} \int_{-\infty}^{\infty}
\frac{ \gamma\alpha^2+\Gamma\omega^2 }{ (\alpha^2+\Delta_1\omega-\omega^2)^2 + \beta^2\omega^2 } d\omega \nonumber\\
&=& \frac{\sqrt{\Gamma}n_{th}}{2\pi} \int_{-\infty}^{\infty}
\frac{1}{\alpha^2} \frac{ \gamma+\Gamma\left( \frac{\omega}{\alpha} \right)^2 }
{ \left( 1+\frac{\Delta_1}{\alpha}\frac{\omega}{\alpha} - \left( \frac{\omega}{\alpha} \right)^2 \right)^2
+ \frac{\beta^2}{\alpha^2}\left( \frac{\omega}{\alpha} \right)^2 } d\omega \nonumber\\
&=& \frac{\sqrt{\Gamma}n_{th}}{2\pi\alpha} \int_{-\infty}^{\infty}
\frac{ \gamma+\Gamma\omega^2 }{ \left( 1+\frac{\Delta_1}{\alpha}\omega-\omega^2 \right)^2
+ \frac{\beta^2}{\alpha^2}\omega^2 } d\omega \nonumber \\
&=& \frac{\sqrt{\Gamma}n_{th}}{2\pi\alpha} \int_{-\infty}^{\infty}
\frac{\gamma+\Gamma\omega^2 }
{ \omega^4 - 2\eta_2\omega^3 + (\eta_2^2-2\eta_1)\omega^2+2\eta_2\omega+1 } d\omega \nonumber\\
&=& \frac{\sqrt{\Gamma}n_{th}}{2\pi\alpha} \int_{-\infty}^{\infty}
\frac{A_1}{\omega-\lambda_1} + \frac{A_2}{\omega-\lambda_2} +
\frac{A_3}{\omega-\lambda_3} + \frac{A_4}{\omega-\lambda_4} d\omega \nonumber\\
\end{eqnarray}
where
\begin{eqnarray}
\eta_1 = 1-\frac{\beta^2}{2\alpha^2}, \quad \eta_2=\frac{\Delta_1}{\alpha},
\end{eqnarray}
\begin{eqnarray}\label{equal time correlation between xi2 and b-eigenvalues}
\lambda_1&=&-i\sqrt{\frac{1-\eta_1}{2}} + \frac{\eta_2}{2} -
\sqrt{ \frac{1+\eta_1}{2} + \frac{\eta_2^2}{4} - i\eta_2\sqrt{\frac{1-\eta_1}{2}} }, \nonumber\\
\lambda_2&=&-i\sqrt{\frac{1-\eta_1}{2}} + \frac{\eta_2}{2} +
\sqrt{ \frac{1+\eta_1}{2} + \frac{\eta_2^2}{4} - i\eta_2\sqrt{\frac{1-\eta_1}{2}} }, \nonumber\\
\lambda_3&=&i\sqrt{\frac{1-\eta_1}{2}} + \frac{\eta_2}{2} -
\sqrt{ \frac{1+\eta_1}{2} + \frac{\eta_2^2}{4} + i\eta_2\sqrt{\frac{1-\eta_1}{2}} }, \nonumber\\
\lambda_4&=&i\sqrt{\frac{1-\eta_1}{2}} + \frac{\eta_2}{2} +
\sqrt{ \frac{1+\eta_1}{2} + \frac{\eta_2^2}{4} + i\eta_2\sqrt{\frac{1-\eta_1}{2}} }, \nonumber\\
\end{eqnarray}
and
\begin{eqnarray}\label{equal time correlation between xi2 and b-coefficients}
A_1&=&\frac{\gamma+\Gamma\lambda_1^2}{(\lambda_1-\lambda_2)(\lambda_1-\lambda_3)(\lambda_1-\lambda_4)}, \nonumber\\
A_2&=&-\frac{ \gamma+\Gamma\lambda_2^2 }{(\lambda_1-\lambda_2)(\lambda_2-\lambda_3)(\lambda_2-\lambda_4)}, \nonumber\\
A_3&=&\frac{ \gamma+\Gamma\lambda_3^2 }{ (\lambda_1-\lambda_3)(\lambda_2-\lambda_3)(\lambda_3-\lambda_4) }, \nonumber\\
A_4&=& -\frac{ \gamma+\Gamma\lambda_4^2 }{(\lambda_1-\lambda_4)(\lambda_2-\lambda_4)(\lambda_3-\lambda_4)}.
\end{eqnarray}
These eigenvalues described in Eq.~(\ref{equal time correlation between xi2 and b-eigenvalues}) can be
rewritten as
\begin{eqnarray}\label{equal time correlation between xi2 and b-eigenvalues2}
\lambda_1&=&\frac{\eta_2}{2}-\sqrt{A}\cos\frac{\theta}{2} + i\left( \sqrt{A}\sin\frac{\theta}{2}-\sqrt{\frac{1-\eta_1}{2}} \right), \nonumber\\
\lambda_2&=&\frac{\eta_2}{2}+\sqrt{A}\cos\frac{\theta}{2} - i\left( \sqrt{A}\sin\frac{\theta}{2}+\sqrt{\frac{1-\eta_1}{2}} \right), \nonumber\\
\lambda_3&=&\frac{\eta_2}{2}-\sqrt{A}\cos\frac{\theta}{2} - i\left( \sqrt{A}\sin\frac{\theta}{2}-\sqrt{\frac{1-\eta_1}{2}} \right),\nonumber\\
\lambda_4&=&\frac{\eta_2}{2}+\sqrt{A}\cos\frac{\theta}{2} + i\left( \sqrt{A}\sin\frac{\theta}{2}+\sqrt{\frac{1-\eta_1}{2}} \right). \nonumber\\
\end{eqnarray}
As $g\gg\Gamma\gg\gamma$ and $\Gamma>\Delta_1$, the above eigenvalues can be approximated to
\begin{eqnarray}\label{equal time correlation between xi2 and b-eigenvalues3}
\lambda_1&\approx&-\sqrt{A}\cos\frac{\theta}{2}-i\sqrt{\frac{1-\eta_1}{2}}, \nonumber\\
\lambda_2&\approx&\sqrt{A}\cos\frac{\theta}{2}-i\sqrt{\frac{1-\eta_1}{2}}, \nonumber\\\
\lambda_3&\approx&-\sqrt{A}\cos\frac{\theta}{2}+i\sqrt{\frac{1-\eta_1}{2}}, \nonumber\\
\lambda_4&\approx&\sqrt{A}\cos\frac{\theta}{2}+i\sqrt{\frac{1-\eta_1}{2}}.
\end{eqnarray}
Substituting Eq.~(\ref{equal time correlation between xi2 and b-eigenvalues3}) into
Eq.~(\ref{equal time correlation between xi2 and b 2}), we have
\begin{eqnarray}
&&\frac{\sqrt{\Gamma}n_{th}}{2\pi} \int_{-\infty}^{\infty}
\frac{ \gamma\alpha^2+\Gamma\omega^2 }{ (\alpha^2+\Delta_1\omega-\omega^2)^2 + \beta^2\omega^2 } d\omega \nonumber\\
&=& \frac{\sqrt{\Gamma}n_{th}}{2\pi\alpha} \int_{-\infty}^{\infty}
\frac{A_1}{\omega-\lambda_1} + \frac{A_2}{\omega-\lambda_2} +
\frac{A_3}{\omega-\lambda_3} + \frac{A_4}{\omega-\lambda_4} d\omega \nonumber\\
&=& \frac{\sqrt{\Gamma}n_{th}}{2\pi\alpha}
i\pi(-A_1-A_2+A_3+A_4) \nonumber\\
&=& \sqrt{\Gamma}n_{th},
\end{eqnarray}
i.e.,
\begin{eqnarray}\label{equal time correlation between xi2 and b 3}
\langle \xi_2^{\dagger}(t)b(t)\rangle + \langle \xi_2^{\dagger}(t)b(t)\rangle^*
\approx \sqrt{\Gamma}n_{th}.
\end{eqnarray}
Similarly, the equal-time correlation $\sqrt{\gamma}\langle \xi_1^{\dagger}b\rangle
+ \sqrt{\Gamma}\langle a^{\dagger}\xi_2\rangle$ (Eq.~(\ref{equal time correlation xi1 b and xi2 a}))
can be calculated as follows
\begin{eqnarray}\label{equal time correlation xi1 b and xi2 a2}
&&\sqrt{\gamma}\langle \xi_1^{\dagger}(t)b(t)\rangle + \sqrt{\Gamma}\langle a^{\dagger}(t)\xi_2(t)\rangle \nonumber\\
&\approx& \frac{g n_{th}\Gamma}{2\pi}
\int_{-\infty}^{\infty} \frac{\frac{\gamma+\Gamma}{2}\omega}
{\left( g^2+\frac{\gamma\Gamma}{4}+\Delta_1\omega-\omega^2 \right)^2
+\left( \frac{\gamma+\Gamma}{2} \right)^2\omega^2} d\omega \nonumber\\
&&+i \frac{g n_{th}\Gamma}{2\pi}
\int_{-\infty}^{\infty}
\frac{ g^2+\frac{\gamma\Gamma}{4} + \Delta_1\omega-\omega^2 }
{\left( g^2+\frac{\gamma\Gamma}{4}+\Delta_1\omega-\omega^2 \right)^2
+\left( \frac{\gamma+\Gamma}{2} \right)^2\omega^2} d\omega \nonumber\\
&\approx& - \frac{g n_{th}\Gamma\beta}{4\alpha^2} \times \nonumber\\
&&\frac{ 2A\sin\theta - \sqrt{2(1-\eta_1)}\eta_2 }
{1+\eta_1^2-\frac{A^2}{2} +2A\cos\theta+\frac{A^2}{2}\cos2\theta-2\eta_1(1+A\cos\theta) } \nonumber\\
&&-i\frac{g n_{th}\Gamma}{4\alpha} \times \nonumber\\
&&\frac{ \sqrt{2(1-\eta_1)}(-1+2A\cos\theta - \eta_1+\frac{\eta_2^2}{2}) - 2A\eta_2\sin\theta }
{1+\eta_1^2-\frac{A^2}{2} +2A\cos\theta+\frac{A^2}{2}\cos2\theta-2\eta_1(1+A\cos\theta) } \nonumber\\
&\approx& 0.
\end{eqnarray}
Finally, by substituting Eqs.~(\ref{equal time correlation between xi1 and a}),
(\ref{equal time correlation between xi2 and b 3}) and (\ref{equal time correlation xi1 b and xi2 a2})
into Eq.~(\ref{Motion equation of second order moments with noise terms}),
the dynamics of the mean photon number $N_a$ and mean phonon number $N_b$ can
be given by
\begin{eqnarray}\label{Reduce motion equation of second order moments with noise terms}
\frac{d N_a}{d t} &=& -\gamma N_a - i g ( \langle a^{\dagger} b\rangle - \langle a^{\dagger} b \rangle^* ), \nonumber\\
\frac{d N_b}{d t} &=& -\Gamma N_b + i g ( \langle a^{\dagger} b\rangle - \langle a^{\dagger} b \rangle^* )
+ \Gamma n_{th}, \nonumber\\
\frac{d \langle a^{\dagger}b\rangle}{d t} &=& - \left (i(\Delta_1-\Delta_2)+\frac{\gamma+\Gamma}{2}\right )
\langle a^{\dagger}b\rangle - i g N_a + i g N_b, \nonumber\\
\end{eqnarray}
and the analytical solution of mean phonon number $N_b(t)$ can be expressed as
\begin{eqnarray}\label{Analytical solution of Nb}
N_b(t) &=& C_1 e^{-\left( \frac{\gamma+\Gamma}{2}+\Gamma_0 \right)t} +
C_2 e^{-\left( \frac{\gamma+\Gamma}{2}-\Gamma_0 \right)t} \nonumber\\
&&+ C_3 e^{-\frac{\gamma+\Gamma}{2}t}\cos(\Omega t) +
C_4 e^{-\frac{\gamma+\Gamma}{2}t}\sin(\Omega t) + N_b^{ss}, \nonumber\\
\end{eqnarray}
where
\begin{eqnarray}
\Omega &=& \frac{1}{2} \sqrt{ \sqrt{  \Upsilon }
+ \left( 8g^2+2\Delta_1^2-\frac{1}{2}(\Gamma-\gamma)^2 \right) }, \\
\Gamma_0&=& \frac{1}{2} \sqrt{ \sqrt{ \Upsilon }
- \left( 8g^2+2\Delta_1^2-\frac{1}{2}(\Gamma-\gamma)^2 \right) }, \\
N_b^{ss}&=& \frac{ 4g^2(\Gamma+\gamma)+\gamma(\Gamma+\gamma)^2+4\gamma\Delta_1^2 }
{ 4g^2(\Gamma+\gamma)+\gamma\Gamma(\Gamma+\gamma)+4\gamma\Gamma\Delta_1^2/(\Gamma+\gamma) }
\cdot \frac{\Gamma}{\Gamma+\gamma}n_{th}, \nonumber\\
\end{eqnarray}
with
\begin{eqnarray}
\Upsilon = \left(8g^2+2\Delta_1^2-\frac{1}{2}(\Gamma-\gamma)^2 \right)^2 + 4(\Gamma-\gamma)^2\Delta_1^2.
\end{eqnarray}
and the coefficients $C_{i}$ can be given by
\begin{eqnarray}
C_1&=& -\frac{ -16g^2\gamma+(\Gamma+\gamma)\left( (\Gamma+\gamma)^2+4\Omega^2 \right) }
{ 16\Gamma_0(\Omega^2+\Gamma_0^2) } n_{th} \nonumber\\
&&-\frac{ + 2\left( 8g^2-(\Gamma+\gamma)^2-4\Omega^2 \right)\Gamma_0 }
{ 16\Gamma_0(\Omega^2+\Gamma_0^2) } n_{th} \nonumber\\
&&+\frac{ \left( (\Gamma+\gamma)^2+4\Omega^2 \right)(\Gamma+\gamma-2\Gamma_0) }
{ 16\Gamma_0(\Omega^2+\Gamma_0^2) } N_b^{ss}, \nonumber\\
C_2&=&-\frac{ 16g^2\gamma - (\Gamma+\gamma)\left( (\Gamma+\gamma)^2+4\Omega^2 \right) }
{ 16\Gamma_0(\Omega^2+\Gamma_0^2) } n_{th} \nonumber\\
&&-\frac{ + 2\left( 8g^2-(\Gamma+\gamma)^2-4\Omega^2 \right)\Gamma_0 }
{ 16\Gamma_0(\Omega^2+\Gamma_0^2) } n_{th} \nonumber\\
&&-\frac{ \left( (\Gamma+\gamma)^2+4\Omega^2 \right)(\Gamma+\gamma+2\Gamma_0) }
{ 16\Gamma_0(\Omega^2+\Gamma_0^2) } N_b^{ss}, \nonumber\\
C_3&=& \frac{ 8g^2+4\Gamma_0^2-(\Gamma+\gamma)^2 }{ 4(\Omega^2+\Gamma_0^2) } n_{th}
+\frac{ (\Gamma+\gamma)^2-4\Gamma_0^2 }{4(\Omega^2+\Gamma_0^2)}N_b^{ss} \nonumber\\
C_4&=& \frac{ 16g^2\gamma+4(\Gamma+\gamma)\Gamma_0^2-(\Gamma+\gamma)^3 }
{ 8\Omega(\Omega^2+\Gamma_0^2) }n_{th} \nonumber\\
&&+\frac{ (\Gamma+\gamma)^2-4\Gamma_0^2 }{8\Omega(\Omega^2+\Gamma_0^2)}
(\Gamma+\gamma)N_{b}^{ss}.
\end{eqnarray}
$\Omega$ is the Rabi oscillation frequency induced by the strong optomechanical interaction and
$N_{b}^{ss}$ is the steady-state cooling limit. Here, we assume that the initial states
of the Stokes field and acoustic field are vacuum state and thermal state, respectively,
i.e., $N_a(t=0)=0, N_b(t=0)=n_{th}$. Eq.~(\ref{Analytical solution of Nb})
can be approximated to
\begin{eqnarray}\label{Reduced analytical solution of Nb}
N_b(t) &\approx& \tilde{C}_1 e^{-\frac{\Gamma+\gamma}{2}t} +
\tilde{C}_2 e^{-\frac{\Gamma+\gamma}{2}t}\cos(\Omega t)
+\tilde{C}_3 e^{-\frac{\Gamma+\gamma}{2}t}\sin(\Omega t)\nonumber\\
&&+ \frac{\Gamma}{\Gamma+\gamma}n_{th},
\end{eqnarray}
where
\begin{eqnarray}
\Omega&\approx&\sqrt{ 4g^2+\Delta_1^2-\frac{(\Gamma-\gamma)^2}{4} }, \nonumber\\
\tilde{C}_1 &\approx& -\frac{ 2(\Gamma-\gamma)g^2-\gamma(\Delta_1^2+\gamma\Gamma) }
{ (\Gamma+\gamma) \Omega^2 } n_{th}, \nonumber\\
\tilde{C}_2 &\approx& \frac{ 2g^2-\gamma(\Gamma+\gamma)/4 }
{ \Omega^2 } n_{th}, \nonumber\\
\tilde{C}_3 &\approx& \frac{ \gamma g - \gamma(\Gamma+\gamma)^2/(16g) }
{ \Omega^2 } n_{th}.
\end{eqnarray}
We show the time evolution of mean phonon number $N_b(t)$ under strong coupling condition
in Figs.~\ref{sup_Fig1} (a) and (b) where blue solid and red dashed curves correspond
to the simulation results and analytical solution described in Eq.~(\ref{Reduced analytical solution of Nb}),
respectively.

\section{Dynamical cooling via pulsed modulation of optomechanical coupling}\label{S3}
In the main text we have mentioned that there exists a state swapping with cycle $T$
between anti-Stokes photons and acoustic phonons in the strong coupling regime, which
includes the swapping cooling and heating processes. We switch on the Brillouin
interaction during the swapping cooling dominated time period to strengthen the
energy transfer from phonons to anti-Stokes photons. When the phonon occupancy
reaches the minimal value, we switch off the Brillouin interaction to suppress
the swapping heating process, i.e., preventing the energy from transferring back
to the anti-Stokes photons. During the switch-off time period, the acoustic mode
is individually driven by the thermal environment. After the light fields pass
through the waveguide and be initialized to the vacuum state, we switch on the Brillouin
interaction again to extract the phonons out of the acoustic field. By modulating
\begin{figure}[h]
\centerline{
\includegraphics[width=8.6 cm,clip]{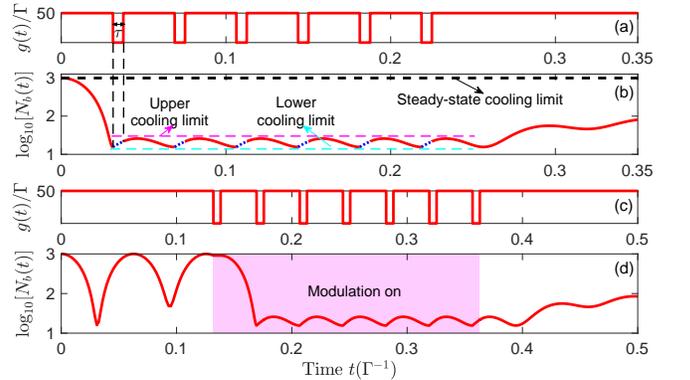}}
\caption{(Color online)
Pulsed modulation scheme of coupling strength $g(t)$ and corresponding
time evolution of phonon occupancy $N_b(t)$. (a) and (b) correspond to
the case where the modulation begins at time $T/2$. The blue dashed
curves in (b) describe the dynamics of phonon occupancy during the pump
switch-off time periods while the switch-off time $\tau=0.1T$. (c) and (d)
correspond to the case where the modulation starts at time $2.1T$.
}\label{sup_Fig2}
\end{figure}
the optomechanical coupling strength to periodically initialize the optical fields,
we can continuously enhance the swapping cooling process while suppress the swapping heating
process and keep the phonon occupation in an instantaneously-state cooling limit
with a small-amplitude fluctuation. This periodical modulation of the coupling
strength can be achieved by a pulsed pump when the Brillouin-active waveguide
is short enough ($L\ll \upsilon_{o}T/2$). We illustrate the pulsed modulation
scheme of the coupling strength and the corresponding time evolution of the
phonon occupancy $N_b(t)$ in Figs.~\ref{sup_Fig2} (a) and (b), respectively,
where $\tau$ represents the switch-off time of pump field. Here, we start the
modulation at the end of the first half Rabi oscillation cycle $t=T/2$ since
the minimum value of $N_b(t)$ is achieved at this time. By substituting $t=T/2=\pi/\Omega$
into Eq.~(\ref{Reduced analytical solution of Nb}), we calculate the instantaneous-state
cooling limit $N_b^{\rm{ins}}$, i.e., the lower cooling limit (red dashed curve) in Fig.~\ref{sup_Fig2} (b),
as follows
\begin{eqnarray}
N_b^{\rm{ins}} &=& N_b(t=\frac{\pi}{\Omega}) \nonumber\\
&\approx& \tilde{C}_1 e^{-\frac{(\Gamma+\gamma)\pi}{2\Omega}}
+\tilde{C}_2 e^{-\frac{(\Gamma+\gamma)\pi}{2\Omega}} \cos(\pi) \nonumber\\
&&+\tilde{C}_3 e^{-\frac{(\Gamma+\gamma)\pi}{2\Omega}} \sin(\pi)+\frac{\Gamma}{\Gamma+\gamma}n_{th} \nonumber\\
&\approx& \frac{ \pi(\Gamma+\gamma)g + \Delta_1^2 - \frac{(\Gamma-\gamma)^2}{4} }
{ 4g^2+\Delta_1^2-\frac{(\Gamma-\gamma)^2}{4} }
\cdot \frac{\Gamma}{\Gamma+\gamma}n_{th} \nonumber\\
&\approx& \frac{\pi\Gamma}{4g}n_{th}.
\end{eqnarray}
In order to evaluate the upper cooling limit $N_b^{\rm{upp}}$, we assume that
the pump switch-off time $\tau$ is small enough that the increase of the phonon
occupancy during the switch-off time can be omitted. Therefore, the analytical
expression of phonon occupancy $N_b(t)$ during time period $[0,\ T/2]$ can be
described by Eq.~(\ref{Reduced analytical solution of Nb}) and during time
period $[T/2,\ T]$ can be approximately given by
\begin{eqnarray}
N_b(t) &\approx& \bar{C}_1 e^{ -\frac{\Gamma+\gamma}{2}(t-\frac{T}{2}) }
+ \bar{C}_2 e^{ -\frac{\Gamma+\gamma}{2}(t-\frac{T}{2}) }\cos(\Omega t) \nonumber\\
&&+ \bar{C}_3 e^{ -\frac{\Gamma+\gamma}{2}(t-\frac{T}{2}) }\sin(\Omega t)
+\frac{\Gamma}{\Gamma+\gamma}n_{th},
\end{eqnarray}
where
\begin{eqnarray}
\bar{C}_1&\approx& \frac{ -8g^2+\pi(\gamma+\Gamma)g }{ 2\Omega^2 }
\cdot \frac{\Gamma}{\Gamma+\gamma}n_{th}, \nonumber\\
\bar{C}_2&\approx& -\frac{ 2\pi(\gamma+\Gamma)g+\Gamma(\Gamma-2\gamma) }{ 4\Omega^2 }
\cdot \frac{\Gamma}{\Gamma+\gamma}n_{th}, \nonumber\\
\bar{C}_3 &\approx& -\frac{ 4g-\pi(\Gamma-\gamma) }{4\Omega^2}
\cdot \Gamma n_{th}, \nonumber\\
\Omega &\approx& \sqrt{4g^2+\Delta_1^2-(\Gamma-\gamma)^2/4}.
\end{eqnarray}
We calculate $N_b(t)$ at time $t=3T/4$ to approximately evaluate $N_b^{\rm{upp}}$,
i.e.,
\begin{eqnarray}
N_b^{\rm{upp}}&\approx& \bar{C}_1 e^{-\frac{\Gamma+\gamma}{2}\frac{\pi}{2\Omega}  }
+ \bar{C}_2 e^{-\frac{\Gamma+\gamma}{2}\frac{\pi}{2\Omega}  } \cos\frac{3\pi}{2} \nonumber\\
&&+ \bar{C}_3 e^{-\frac{\Gamma+\gamma}{2}\frac{\pi}{2\Omega}  } \sin\frac{3\pi}{2}
+ \frac{\Gamma}{\Gamma+\gamma}n_{th} \nonumber\\
&\approx& \frac{(1+\pi)\Gamma}{4g}n_{th}.
\end{eqnarray}

\begin{figure}[h]
\centerline{
\includegraphics[width=8.8 cm,clip]{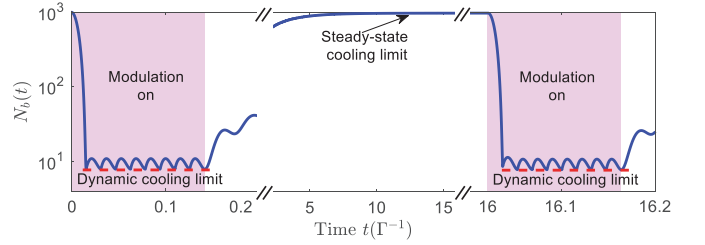}}
\caption{(Color online) The pulsed modulation of the coupling strength
is switchable where the regions with light purple color correspond to
the modulation turned on.
}\label{sup_Fig3}
\end{figure}

In Fig.~\ref{sup_Fig2} (b), we apply the modulation of the coupling strength at the
end of first half Rabi oscillation cycle since the  phonon occupancy reaches the
minimum value at this time. Actually, we can turn on the modulation at any time to
cool the phonons. We show the time evolution of the phonon occupancy while the modulation
is turned on at time $t=2.1T$ in Fig.~\ref{sup_Fig2} (d), where Fig.~\ref{sup_Fig2} (c)
denotes the modulation scheme. The acoustic mode reaches the instantaneous-state cooling
limit when the modulation is turned on and transits back to the steady-state cooling
limit while the modulation is turned off. It means that this dynamic cooling scheme
with pulsed modulation of the optomechanical coupling intensity is switchable, as shown
in Fig.~\ref{sup_Fig3}.

\section{Continuum optomechanical cooling via Brillouin interaction}\label{S4}
\begin{figure}[h]
\centerline{
\includegraphics[width=8.4 cm,clip]{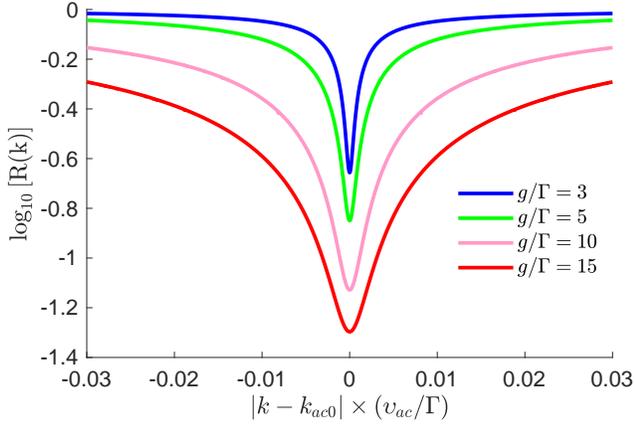}}
\caption{(Color online) Continuum optomechanical cooling versus wavenumber $k$ for
different strong coupling strength $g/\Gamma = 3, 5, 10, 15$.
}\label{sup_Fig4}
\end{figure}

In cavity optomechanical systems, as the optical and the mechanical modes which
cause the optomechanical interaction are discrete modes, the sideband cooling method
can only generate a net cooling effect on a single mechanical mode~\cite{Teufel2,Chan}.
However, in continuum optomechanical systems, for example, Brillouin-active waveguides,
the optomechanical interaction involves the acoustic field with a continuous band of
accessible modes, which enables the continuum optomechanical cooling~\cite{Otterstrom,Chen}.
In fact, as described in Eq.~(\ref{Envelope operators with continuous wavenumber}),
the envelope operator $b_{ac}$ of the acoustic filed associated in the Brillouin interaction
is an operator with continuous wavenumber $k$, which is peaked around a carrier wave vector
$k_{ac0}$ and enables the acoustic field to evolve in space. Thus, in the momentum space,
if we consider an undepleted pump, we can obtain motion equations as described in
Eq.~(\ref{Motion equation of linearized Brillouin with reduced variables}) for each
specific wavenumber $k$ and then achieve the dynamics of the mean phonon number $N_b(k,t)=\langle b^{\dagger}(k,t)b(k,t)\rangle$
for the $k$-th acoustic mode, as described in Eq.~(\ref{Reduce motion equation of second order moments with noise terms}).
Finally, by periodically modulating the Brillouin interaction via a pulsed pump in
a short enough waveguide, the optomechanical cooling with continuum acoustic modes
can become accessible. We show the simulation results of the continuum optomechanical
cooling under the strong coupling regime in Fig.~\ref{sup_Fig4}. It can be clearly seen
that the largest cooling ratio $R$ is achieved at point $k=k_{ac0}$, i.e., the phase-matching
point. The cooling ratio decreases with $|k-k_{ac0}|$ because of the breaking of phase-matching condition.
When $|k-k_{ac0}|$ is too large that the phase-matching condition of the Brillouin
scattering is completely unsatisfied, the phonon cooling will disappear. It demonstrates
that the continuum optomechanical cooling with a broad band can be achieved in
Brillouin-active waveguides by using our modulation scheme.

\section{Dynamical cooling generated by forward intermodal Brillouin scattering}\label{S5}
\begin{figure}[h]
\centerline{
\includegraphics[width=8.4 cm,clip]{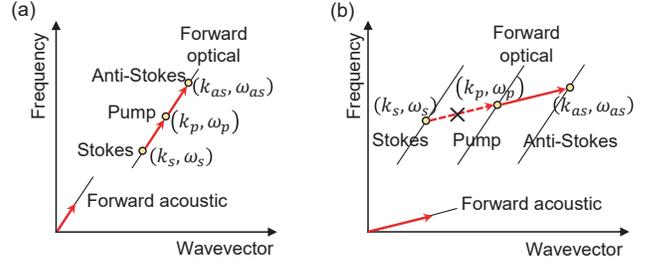}}
\caption{(Color online) Sketch of the dispersion diagram of forward
Brillouin interaction for both intra-modal scattering (a) and
inter-modal scattering (b).
}\label{sup_Fig5}
\end{figure}

In this section, we apply the modulation scheme of the optomechanical coupling intensity
to the Brillouin cooling generated by the forward Brillouin anti-Stokes scattering in continuum
optomechanical systems while optical dissipation exceeds mechanical dissipation~\cite{Otterstrom}.
In order to achieve the Brillouin cooling, a critical requirement is the suppression of the
Stokes scattering process. For the intra-modal Brillouin forward scattering, the Stokes process
can not be suppressed since the Stokes and anti-Stokes waves interact through the same
phonon mode~\cite{Kharel}, as shown in Fig.~\ref{sup_Fig5} (a). However, the dispersive
symmetry between the Stokes and anti-Stokes processes can be broken for the inter-modal
Brillouin forward scattering~\cite{Chen,Otterstrom} when only the anti-Stokes process is
engineered to satisfy the phase-matching condition, as shown in Fig.~\ref{sup_Fig5} (b).
Here, we use the inter-modal forward anti-Stokes Brillouin scattering to study the
phonon cooling.

Actually, similar to the backward Brillouin anti-Stokes scattering, there exists a Rabi
oscillation of the phonon occupancy in the strong coupling regime for the forward Brillouin
anti-Stokes scattering, which causes the reversible state swapping between phonons and
anti-Stokes photons, i.e., swapping cooling and heating processes. If the
size of the active-Brillouin waveguide is short enough which enables the Brillouin
optomechanical interaction can be switched on and off by a pulsed pump, a dynamic
cooling limit can be achieved by modulating the Brillouin optomechanical interaction
to enhance the swapping cooling process while suppressing the heating process, which
breaks the steady-state cooling limit.

\begin{figure}[h]
\centerline{
\includegraphics[width=8.6 cm,clip]{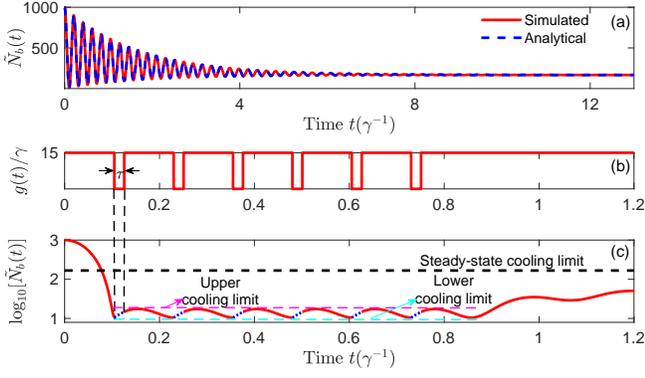}}
\caption{(Color online) (a) The time evolution of $\tilde{N}_b (t)$ in the strong
coupling regime for $g/\gamma=15$. Modulation scheme of $g(t)$ with
the switch-off time $\tau=0.1T$ (a) and the corresponding time evolution of the
phonon occupancy (b). Other parameters are $\frac{\Gamma}{\gamma} = 0.2$, $\frac{\Delta_1}{\gamma}
=0.05$,  $\frac{\Delta_2}{\gamma}=0.05\times 10^{-4}$, and $n_{th}=1000$.
}\label{sup_Fig6}
\end{figure}

\begin{figure}[h]
\centerline{
\includegraphics[width=8.6 cm,clip]{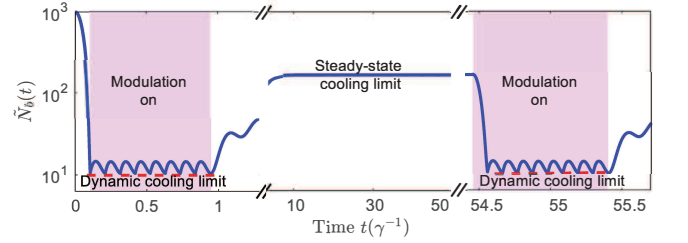}}
\caption{(Color online) The pulsed modulation is switchable.
}\label{sup_Fig7-2}
\end{figure}

Considering an un-depleted pump, the triply resonant Brillouin interaction can be
reduced to an optomechanical interaction between anti-Stokes photons and acoustic
phonons with a pump-enhanced coupling strength $g$. The dynamics of this reduced
optomechanical interaction in the momentum space can be expressed as
\begin{eqnarray}\label{Motion equation of linearized Brillouin with reduced variables-forward}
\frac{d\tilde{a}}{dt}&=& \left( -\frac{\gamma}{2}+i\Delta_1 \right)\tilde{a} - i g
\tilde{b} + \sqrt{\gamma}\tilde{\xi}_1, \nonumber\\
\frac{d\tilde{b}}{dt}&=& \left(  -\frac{\Gamma}{2}+i\Delta_2 \right)\tilde{b} - i g
\tilde{a} + \sqrt{\Gamma}\tilde{\xi}_2,
\end{eqnarray}
where $\tilde{a}$ ($\tilde{b}$) denotes the annihilation operator of the anti-Stokes
photons (acoustic phonons) corresponding to the $k$-th photon (phonon) mode.
$\gamma$ ($\Gamma$) represents the optical (acoustic) dissipation. $\Delta_1=\upsilon_{o}k$
and $\Delta_2=\upsilon_{ac}k$ indicate the frequency shifts induced by the wavenumber
$k$ for the anti-Stokes photons and acoustic phonons, respectively, where $\upsilon_{o}$
($\upsilon_{ac}$) is the group velocity of the photons (phonons). $\tilde{\xi}_1$ and
$\tilde{\xi}_2$ are the Langevin noises of the anti-Stokes field and the acoustic field,\
which obeys relationships
\begin{eqnarray}
\langle \xi_1(t)\rangle &=& 0, \nonumber\\
\langle \xi_1^{\dagger}(t_1)\xi_1(t_2) \rangle &=& 0, \nonumber\\
\langle \xi_1(t_1) \xi_1^{\dagger}(t_1) \rangle &=& \delta(t_1-t_2), \nonumber\\
\langle \xi_2(t)\rangle &=& 0, \nonumber\\
\langle \xi_2^{\dagger}(t_1) \xi_2(t_2) &=& n_{th}\delta(t_1-t_2), \nonumber\\
\langle \xi_2(t_1) \xi_2^{\dagger}(t_2) &=& (1+n_{th})\delta(t_1-t_2),
\end{eqnarray}
where $n_{th}$ correspond the thermal phonon occupation. Thus the differential
equations for the second-order moments $\tilde{N}_a=\langle \tilde{a}^{\dagger}\tilde{a}\rangle$,
$\tilde{N}_b=\langle \tilde{b}^{\dagger}\tilde{b}\rangle$, $\langle \tilde{a}^{\dagger}\tilde{b}\rangle$
can be given by
\begin{eqnarray}\label{Motion equation of second order moments with noise terms-forward}
\frac{d\tilde{N}_a}{dt}&=& -\gamma\tilde{N}_a - i g\left[ \langle \tilde{a}^{\dagger}\tilde{b} \rangle
-\langle \tilde{a}^{\dagger}\tilde{b} \rangle^* \right], \nonumber\\
\frac{d\tilde{N}_b}{dt}&=& -\Gamma \tilde{N}_b + i g\left[ \langle \tilde{a}^{\dagger}\tilde{b} \rangle
-\langle \tilde{a}^{\dagger}\tilde{b} \rangle^* \right] \nonumber\\
&&+\frac{\Gamma n_{th}}{2\pi}
\int_{-\infty}^{\infty} \frac{ \gamma(g^2+\frac{\gamma\Gamma}{4})+\Gamma(\omega+\Delta_1)^2 }
{ \tilde{\Xi}_1 } d\omega, \nonumber\\
\frac{d\langle \tilde{a}^{\dagger}\tilde{b}\rangle}{dt}&=& - \left[ i(\Delta_1-\Delta_2)+\frac{\gamma+\Gamma}{2} \right]
\langle \tilde{a}^{\dagger}\tilde{b}\rangle - i g \tilde{N}_a + i g \tilde{N}_b \nonumber\\
&&+ \frac{i g \Gamma n_{th}}{2\pi}\int_{-\infty}^{\infty} \frac{1}
{ \tilde{\Xi}_2 } d\omega,
\end{eqnarray}
with
\begin{eqnarray}
\tilde{\Xi}_1 &=& \left[ g^2+\frac{\gamma\Gamma}{4}-(\omega+\Delta_1)(\omega+\Delta_2) \right]^2 \nonumber\\
&&+ \left[ \frac{\gamma+\Gamma}{2}\omega+\frac{\gamma\Delta_2+\Gamma\Delta_1}{2} \right]^2, \nonumber\\
\tilde{\Xi}_2 &=& g^2+\frac{\gamma\Gamma}{4}-(\omega+\Delta_1)(\omega+\Delta_2)  \nonumber\\
&&+ i\left[ \frac{\gamma+\Gamma}{2}\omega+\frac{\gamma\Delta_2+\Gamma\Delta_1}{2} \right],
\end{eqnarray}
where $\tilde{N}_a$ and $\tilde{N}_b$ correspond to the mean photon and phonon numbers.
We consider the strong coupling regime ($g\gg\gamma,\Gamma$) and assume that
the optical dissipation is far larger than the mechanical dissipation ($\gamma\gg\Gamma$)
and the wavenumber induced frequency shifts are within the linewidth of the acoustic mode ($\Delta_{1,2}<\Gamma$).
In addition, for the forward Brillouin scattering in a typical Brillouin-active waveguide,
the velocity of the optical fields is greatly lager than the velocity of the acoustic field
which leads to $\Delta_1\gg\Delta_2$ for $k\neq0$. Thus the integral terms in
Eq.~(\ref{Motion equation of second order moments with noise terms-forward}) can be
approximated to
\begin{eqnarray}
\int_{-\infty}^{\infty} \frac{ \gamma(g^2+\frac{\gamma\Gamma}{4})+\Gamma(\omega+\Delta_1)^2 }
{ \tilde{\Xi}_1 } d\omega
&\approx& 2\pi, \nonumber\\
\int_{-\infty}^{\infty} \frac{1}
{ \tilde{\Xi}_2 } d\omega
&\approx& 0.
\end{eqnarray}
Substituting the above equations into Eq.~(\ref{Motion equation of second order moments with noise terms-forward}),
the analytical solution of the mean phonon number in the strong coupling regime can be expressed as
\begin{eqnarray}
\tilde{N}_b &\approx& \tilde{C}_1 e^{-\frac{\gamma+\Gamma}{2}t} + \tilde{C}_2 e^{-\frac{\gamma+\Gamma}{2}t}
\cos(\Omega t)  \nonumber\\
&&+ \tilde{C}_3 e^{-\frac{\gamma+\Gamma}{2}t} \sin(\Omega t) + \tilde{N}_b^{ss},
\end{eqnarray}
with
\begin{eqnarray}
\tilde{C}_1 &\approx& \frac{ 2g^2+\Delta_1^2-\frac{(\Gamma-\gamma)^2}{4} + \frac{(\gamma+\Gamma)^2}{4} }
{ \Omega^2 } n_{th} \nonumber\\
&&- \frac{ 4g^2 + \Delta_1^2 - \frac{(\Gamma-\gamma)^2}{4} + \frac{(\Gamma+\gamma)^2}{4} }
{ \Omega^2 } \tilde{N}_b^{ss}, \nonumber\\
\tilde{C}_2 &\approx& \frac{ 2g^2 -\frac{(\Gamma+\gamma)^2}{4} }
{ \Omega^2 } n_{th}
+ \frac{ \frac{(\Gamma+\gamma)^2}{4} }
{ \Omega^2 } \tilde{N}_b^{ss}, \nonumber\\
\tilde{C}_3 &\approx& \frac{ g\gamma - \frac{(\gamma+\Gamma)^3}{16g}}
{ \Omega^2 } n_{th}
+ \frac{ \frac{(\gamma+\Gamma)^3}{16g} }
{ \Omega^2 } \tilde{N}_b^{ss}, \nonumber\\
\tilde{N}_b^{ss} &\approx& \frac{ 4g^2+\gamma(\gamma+\Gamma) }{ 4g^2+\gamma\Gamma }\cdot
\frac{\Gamma}{\Gamma+\gamma}n_{th}, \nonumber\\
\Omega&\approx& \sqrt{ 4g^2+\Delta_1^2 - \frac{(\Gamma-\gamma)^2}{4}},
\end{eqnarray}
where $\tilde{N}_b^{ss}$ denotes the steady-state cooling limit. We show the simulation
results and the analytical solution of the time evolution for the phonon occupancy
$\tilde{N}_b(t)$ in the strong coupling regime in Fig.~\ref{sup_Fig6} (a). It can be
clearly seen that the strong optomechanical interaction enables the phonon occupancy
to exhibit a Rabi oscillation with cycle $T=2\pi/\Omega$. The minimum value of the
phonon occupancy can be achieved at the end of the first half Rabi oscillation cycle
$t_{\rm{min}}=T/2$, which is much smaller than the steady-state cooling limit. Therefore,
we switch off the pump at time $t_{\rm{min}}$ to halt the reversible Rabi oscillation
and suppress the swapping heating process. After optical fields pass through the waveguide
and are initialized to the vacuum state, we switch on the pump again to generate the swapping
cooling process for absorbing phonons. We illustrate the modulation scheme and the
corresponding time evolution of the phonon occupancy in Figs.~\ref{sup_Fig6} (b) and (c),
respectively. By periodically initializing optical fields through a pulsed pump,
the phonon occupancy can be continuously suppressed to an instantaneous-state cooling
limit with a small-amplitude fluctuation, which breaks the steady-state cooling limit,
as shown in Fig.~\ref{sup_Fig6} (c) where $\tau$ is the switch-off time. The
instantaneous-state cooling limit, i.e., the lower cooling limit in Fig.~\ref{sup_Fig6} (c),
can be evaluated at time $t_{\rm{min}}$ and approximately expressed as $\pi\Gamma n_{th}/(4g)$.
In order to evaluate the upper cooling limit, we assume that switch-off time $\tau$ is
small enough that the increase of the phonon occupancy during the switch-off time can
be ignored. Then the analytical expression of $\tilde{N}_b(t)$ during time period
$[T/2, T]$ can be given by
\begin{eqnarray}
\tilde{N}_b (t) &\approx& \bar{C}_1 e^{-\frac{\gamma+\Gamma}{2}(t-T/2)}
+ \bar{C}_2 e^{-\frac{\gamma+\Gamma}{2}(t-T/2)} \cos(\Omega t) \nonumber\\
&&+ \bar{C}_3 e^{-\frac{\gamma+\Gamma}{2}(t-T/2)} \sin(\Omega t) + \bar{N}_b^{ss},
\end{eqnarray}
where
\begin{eqnarray}
\bar{C}_1 &\approx& \frac{ -16g^2+2\pi(\gamma+\Gamma)g }{4\Omega^2}
\cdot\frac{\Gamma}{\Gamma+\gamma}n_{th}, \nonumber\\
\bar{C}_2 &\approx& -\frac{ 2\pi(\gamma+\Gamma)g-(3\gamma^2+2\gamma\Gamma-\Gamma^2) }{4\Omega^2}
\cdot\frac{\Gamma}{\Gamma+\gamma}n_{th}, \nonumber\\
\bar{C}_3 &\approx& - \frac{4g+\pi(\gamma-\Gamma)}{ 4\Omega^2 } \Gamma n_{th}, \nonumber\\
\bar{N}_b^{ss}&\approx& \frac{ 4g^2+\gamma(\gamma+\Gamma) }{4g^2+\gamma\Gamma}
\cdot\frac{\Gamma}{\Gamma+\gamma}n_{th}, \nonumber\\
\Omega &\approx& \sqrt{ 4g^2+\Delta_1^2-\frac{(\gamma-\Gamma)^2}{4} }.
\end{eqnarray}
The upper cooling limit in Fig.~\ref{sup_Fig6} (c) can be approximately calculated
around time $t=3T/4$
\begin{eqnarray}
\bar{N}_b^{\rm{upp}} &\approx& \bar{C}_1 e^{-\frac{\gamma+\Gamma}{2}\frac{\pi}{2\Omega}}
-\bar{C}_3 e^{-\frac{\gamma+\Gamma}{2}\frac{\pi}{2\Omega}} + \bar{N}_b^{ss} \nonumber\\
&\approx& \frac{(1+\pi)\Gamma}{4g}n_{th}.
\end{eqnarray}
Furthermore, this modulation scheme is switchable by turning on or off the modulation, as shown in Fig.~\ref{sup_Fig7-2}.

\end{document}